\documentclass[preprint]{aastex}
\usepackage{graphicx}

\newcommand{\numax}{$\nu_{max}$}

\begin{document}

\title{High frequency quasi-periodic oscillations in the black hole
X-ray transient XTE J1650--500}

\author{Jeroen Homan\altaffilmark{1}, Marc
Klein-Wolt\altaffilmark{2}, Sabrina Rossi\altaffilmark{1}, Jon M.\
Miller\altaffilmark{3,6}, Rudy Wijnands\altaffilmark{4}, Tomaso
Belloni\altaffilmark{1}, Michiel van der Klis\altaffilmark{2}, \&
Walter H.\ G.\ Lewin\altaffilmark{5}}

\altaffiltext{1}{INAF - Osservatorio Astronomico di Brera, Via E.\
Bianchi 46, 23807 Merate (LC); homan@merate.mi.astro.it, 
srossi@merate.mi.astro.it, belloni@merate.mi.astro.it }
\altaffiltext{2}{Astronomical Institute `Anton Pannekoek', University
of Amsterdam, and Center for High Energy Astrophysics, Kruislaan 403,
1098 SJ, Amsterdam, the Netherlands; klein@astro.uva.nl,
michiel@astro.uva.nl} \altaffiltext{3}{Harvard-Smithsonian Center for
Astrophysics, 60 Garden Street, Cambridge, MA 02138;
jmmiller@head-cfa.harvard.edu} \altaffiltext{4}{School of Physics and
Astronomy, University of St  Andrews, North Haugh, St Andrews Fife,
KY16 9SS, Scotland, UK; radw@st-andrews.ac.uk}
\altaffiltext{5}{Center for Space Research and Department of Physics,
Massachusetts Insitute of Technology, Cambridge, MA 02139-4307;
lewin@space.mit.edu} \altaffiltext{6}{National Science Foundation
Astronomy \& Astrophysics Fellow}

\begin{abstract} We report the detection of high frequency
variability in the black hole X-ray transient XTE J1650--500. A
quasi-periodic oscillation (QPO) was found at 250 Hz during a
transition from the hard to the soft state. We also detected less
coherent variability around 50 Hz, that disappeared when the 250 Hz
QPO showed up. There are indications that when the energy spectrum
hardened the QPO frequency increased from $\sim$110 Hz to $\sim$270
Hz, although the observed frequencies are also consistent with being
1:2:3 harmonics of each other. Interpreting the 250 Hz as the orbital
frequency at the innermost stable orbit around a Schwarzschild black
hole leads to a mass estimate of 8.2 $M_\odot$. The spectral results
by Miller et al.\ (2002, ApJ, 570, L69), which suggest considerable
black hole spin, would imply a higher mass.    \end{abstract}


\section{Introduction}\label{sec:intro}

The majority of the known stellar mass black hole candidates (BHCs)
are transient low-mass X-ray binaries. These systems remain in
quiescence for periods of months to tens of years and occasionally
show outbursts (X-ray novae) which are the result of a sudden
increase in the mass accretion rate that generally lasts for a few
months. During an outburst  various spectral and variability
properties are observed which seem to a great extent, but not
uniquely, to be determined by the mass accretion rate \citep[see
e.g.][]{tale1995,howiva2001,no2002}. Although the appearance of black
hole X-ray binaries is often categorized into 4 to 5 canonical
states, they in fact appear to go through a continuous range of
states whose properties are determined by the interplay between the
two dominating spectral components: a soft thermal disk component and
a hard non-thermal power law component attributed to a Comptonizing
medium \citep{sutr1979} whose origin remains unclear but is often
thought to be an outflow and/or corona surrounding the inner
accretion disk. 

The variability properties of BHCs are strongly related to the
relative contribution of the disk and the non-thermal components.
When the latter dominates (hard state), the power spectra of BHCs
show a strong band limited noise component on which often are
superposed one or more quasi-periodic oscillations (QPOs). When  the
contribution of the non-thermal component to the spectrum decreases
(intermediate state), the noise becomes weaker and the QPOs become
more prominent, while their frequencies increase from typically 0.1
to 10 Hz. When the thermal component dominates the spectrum (soft
state) variability is very weak. 

In addition to low frequency QPOs (LF QPOs; $\sim$0.1--10 Hz)  five
BHC systems have also shown high frequency QPOs (HF QPOs) with
frequencies of 40 Hz and higher \citep[for a recent overview
see][]{remumc2002b}.  They are mainly observed when neither the
thermal nor the non-thermal component dominates the spectrum. Unlike
the kHz QPOs that are observed in the neutron star low-mass X-ray
binaries \citep{va2000} and whose frequencies are variable and
correlated to those of QPOs at lower frequency, our view on HF QPOs
in BHCs is less complete and has evolved considerably during the past
six years. Currently, three of the five sources have shown pairs HF
QPOs, sometimes simultaneously detected: GRO 1655--40 shows a pair at
300 and 450 Hz \citetext{\citealp{st2001a}; see also
\citealp{remomc1999}}, GRS 1915+105  at 41 and 67 Hz 
\citetext{\citealp{st2001b}; see also \citealp{moregr1997}} and
another at 164 and 328 Hz \citep[not detected simultaneously with the
other pair]{remumc2002b}, and XTE J1550--564 at 184 and 272 Hz
\citep{miwiho2001,remumc2002a}. Note that in the case of XTE
J1550--564 there is also evidence for a peak at 92 Hz
\citep{remumc2002a} and that single peaks at intermediate frequencies
are also found \citep{howiva2001,remumc2002a}. The frequencies of
the  102--284 Hz peaks in XTE J1550--564 reported by
\citet{howiva2001} seemed to correlate  with the frequency of the LF
QPOs, although \citet{resomu2002} dispute this. In XTE J1859+226
\citep{cushha2000,ma2001} and 4U 1630--47 \citep{remo1999} only
single peaks are detected that are consistent (at a $3\sigma$ level)
with having a constant frequency. 

Although several detailed models had already been proposed to explain
the frequencies of the HF QPOs (some of which will be discussed in
Section \ref{sec:disc}), recently much interest has been sparked by
the apparent harmonic  ratios of the frequencies in the HF QPO pairs
(3:5 and 1:2 in GRS 1915+105, 2:3 in GRO J1650--40, and 1:2:3 XTE
J1550--564), which, according to \citet{abkl2001},  might be
suggestive of resonances between orbital and epicyclic motion in the
Kerr metric. In any case, the frequencies of the HF QPOs strongly
suggest that they find their origin in the inner parts of the
accretion flow, close to the innermost stable orbit (ISCO). When
eventually the mechanism behind these oscillations is understood,
their properties can be used to  study the effects of General
Relativity  and constrain the mass and spin of the black hole.
Already now such constraints can be made; for example, assuming that
the highest observed frequencies are caused by Keplerian motion at
the ISCO, for at least two of the above sources, GRO J1655--40 and
XTE J1550--564, the derived values of the masses of the black holes
\citep[5.5--7.9 $M_\odot$ and 9.7--11.6,][]{shvaca1999,orgrva2002}
and the highest observed frequencies imply that the black holes {\it
must} be spinning. The models proposed thus far to explain the
frequencies of the HF QPOs have all in common that their
interpretation of the highest observed frequency requires even larger
spin values than the ones based on the above simple assumption.

In this paper we report our search for high frequency QPOs in the
X-ray transient XTE J1650--500. This source was first detected with
the Rossi X-ray Timing Explorer (RXTE) on September 5 2001
\citep{re2001}. Follow up observations
\citep{msswsm2001,resu2001,wimile2001} revealed a hard X-ray
spectrum, strong variability and QPOs around a few Hz, marking the
source as a black hole candidate. Optical and radio counterparts were
identified by \citet{cakigi2001} and \citet{grudmi2001}. Optical
follow-up observations performed when the source was close to
quiescence, in October 2002, revealed an orbital period of 0.212 days
and a mass function which requires an inclination $<40^\circ$ for the
black hole mass to be larger than 3 $M_\odot$ \citep{sazuca2002}.
XMM-Newton observations performed near the peak of the outburst
revealed the presence of a broad, skewed Fe K$_\alpha$ emission line
that suggests the compact object in this system may be a Kerr black
hole \citep{mifawi2002}. \citet{katoro2002} studied the variability
properties during a $\sim$20 day interval centered around a
transition from the soft to the hard state that occurred on 2001
November 18. In the power spectra they found LF QPOs between 4 and 9
Hz, but also broad bumps around 25 and 80 Hz. No HF QPOs were found,
however. During the later stages of the outburst a 14-day period was
found in the ASM light curves, which is probably the disk precession
period \citep{tokaco2002}.  In this paper we report the detection of
high frequency variability in this source and in particular the
detection of a 250 Hz QPO. A more detailed spectral and variability
study of the outburst will be presented elsewhere.

\section{Observations and analysis}\label{sec:obs} 

The data used in this paper were taken with the proportional counter
array \citep[PCA,][]{jaswgi1996} onboard RXTE \citep{brrosw1993}. The
observations of XTE J1650--500 were performed between 2001 September
6 and 2001 November 22. A total of 85 pointed observations were made,
amounting to $\sim$185 ks. The data used in our analysis were in the
following modes: {\tt Standard 2}, which has a time resolution of 16
s and covers the 2--60 keV PCA effective energy range with 129 energy
channels, and {\tt E\_125us\_64M\_0\_1s}/{\tt E\_64us\_64M\_0\_1s}
(Event modes), which provide a time resolution of $2^{-13}$ s or
$2^{-14}$ s in 64 energy channels covering the same energy range as
the {\tt Standard 2} mode. 

The {\tt Standard 2} data were used to create a light curve,  a
hardness-intensity diagram (HID) and a color-color diagram (CD). Only
data from proportional counter units (PCUs) 0 and 2 were used since
these were the only two (of the five) that were always operational
during our observations. The observations were background subtracted,
but no dead time corrections were applied ($<$3.6\%). Two colors were
defined; a soft color, which is the ratio of counts in the 5.8--10.0
keV and 2.5--5.8 keV bands ({\tt Standard 2} [i.e.\ 0--128] channels
10--19 and 2--9), and a hard color, which is the ratio of counts in
the 13.8--18.4 keV and 2.5--5.8 keV bands (channels 29--39 and 2--9).
The 'intensity' used for the HID was the count rate in the 2.5--20.5
keV band (channels 2--44).

Fast Fourier transforms (FFTs) of the Event mode data (from all the
active PCUs) were performed to create power spectra with a frequency
range of 0.0625--4096 Hz in the 7.5--22.2 keV band (`absolute' [i.e.\
0--254] channels 15-49). This energy range was chosen since high
frequency variability in black hole LMXBs is in general most
significantly detected in a similar energy range - an additional
advantage of this energy range is that it does not introduce the
instrumental broad high frequency features found by Klein-Wolt, Homan
\& van der Klis (in prep.) in the lowest energy channels.  The data
were not background subtracted and no dead time corrections were
applied prior to the FFT; the effects of dead time were accounted for
by our power spectral fit function (see below). Power spectra were
averaged based on time, colors or count rate, and normalized
according to the recipe described in \citet{va1995b}. Prior to
fitting the power spectra the dead time modified Poisson level was
subtracted using the method of Klein-Wolt, Homan \& van der Klis (in
prep.), which uses the analytical function from \citet{zh1995} and
\citet{zhjasw1995} as a basis. They found that this function often
fails to describe the high frequency part of the power spectrum where
no contribution from the source is expected and,  instead of
improving the fit at high frequencies by varying the parameters of
this function (which leads to unacceptable values for e.g. the
instrumental dead time and VLE window), the Zhang function is scaled
to provide the best fit to the power spectrum in the 1000--4000 Hz
range. 

Following a current trend initiated by \citet{olbabo1998}
\citet{no2000} we fit the resulting power spectra with a sum of
Lorentzians, each given by $P(\nu)=(r\Delta/\pi)[\Delta^2 +
(\nu-\nu_0)^2]^{-1}$, where $\nu_0$ is the centroid frequency,
$\Delta$ the half-width-at-half-maximum, and $r$ the integrated
fractional rms (from $-\infty$ to $\infty$). Instead of $\nu_0$ and
$\Delta$ we will quote the frequency at which the Lorentzian attains
its maximum in $\nu P(\nu)$, $\nu_{max}$, and the quality factor, Q,
where $\nu_{max}=\nu_0 (1 + 1/4Q^2)^{1/2}$ and $Q=\nu_0/2\Delta$
\citep{bepsva2002}. The fractional rms amplitudes quoted in this
paper are the integrated power between 0 and $\infty$. Errors on fit
parameters were determined using $\Delta\chi^2=1$. Upper limits on
the strength of the Lorentzians were determined  by fixing the Q
and/or \numax, but not the rms amplitude, to  values similar to those
obtained in another power spectrum and using $\Delta\chi^2=2.71$
(95\% confidence).

\begin{figure}[t] 
\centerline{\includegraphics[width=12cm]{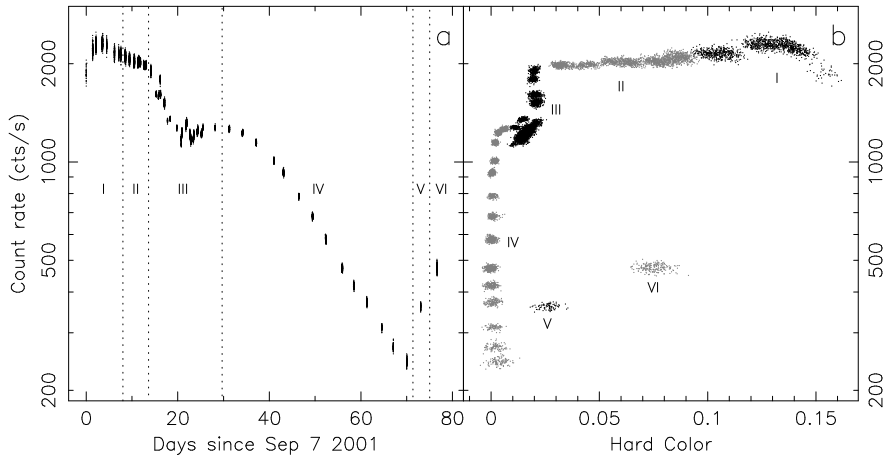}}
\caption{The light curve (a) and hardness-intensity diagram (b) of
the outburst of XTE J1650--500. Count rates are in the 2.5--20.5 keV
band, and the hard color in panel (b) is the ratio of counts in the
13.8--18.4 keV and 2.5--5.8 keV bands. Each points represent a 16 s
interval. The 6 groups for the power spectral analysis are indicated
by Roman numerals I--VI and are separated by vertical lines in panel
(a) and plotted alternately in black and gray in panel (b). The
general motion of the source in the HID is
anti-clockwise.\label{fig:lc-hid}} \end{figure}

\section{Results}\label{sec:res}   

Figure \ref{fig:lc-hid} shows the 2.5--20.5 keV light curve and HID
of the outburst of XTE J1650--500. The first few observations were
done during the final stages of the rise, when the source was very
hard. After reaching a peak on September 10 (day 3) a more or less
exponential decline followed, lasting for $\sim$8 days, during which
the spectral hardness changed considerably. The period between
September 25 (day 18) and October 7 (day 30), which clearly shows up
as a dent in the light curve, marks a transition from the hard to the
soft state  - during this period the day to day variations in count
rate and hardness were more erratic than before and after, while both
show an overall decrease. After this transition the changes were 
smoother, and around October 17 (day 40) the count rate started to
decline exponentially again. A minimum in the hardness was reached
around November 1 (day 55). After November 16 (day 70) the source
showed a remarkable hardening accompanied by a strong increase in
count rate. Part of the decline ($\sim$20 days centered on our last
observation) is discussed in \citet{katoro2002}.

\begin{table}[b!]
\vspace{1cm}
{\footnotesize
\begin{tabular}{cccccccccc}
\hline
\hline
  &  \multicolumn{3}{c}{Noise} & \multicolumn{3}{c}{LF QPO} & \multicolumn{3}{c}{HF QPO} \\
Sel. & \numax\ (Hz) & Q & rms (\%) & \numax\ (Hz) & Q & rms (\%) & \numax\ (Hz) & Q & rms (\%)\\
\hline
I   & 0.51$\pm$0.04	  & 0.11$\pm$0.05 & 12.6$\pm$0.5 & 1.3--9.6    & 0.3--1.5    & 5.8--11.4 & 53$\pm$2    & 0.24$\pm$0.05       & 9.4$\pm$0.4 \\
II  & 0.80$\pm$0.02	  & 0.33$\pm$0.02 & 8.5$\pm$0.2  & 1.7--8.2    & 0.4--3.0    & 2.6--14.2 & 55$\pm$5    & 0.49$\pm$0.12       & 6.9$\pm$0.7 \\
    & ---		  & --- 	  & --- 	 & ---         & ---	     & ---	 & 139$\pm$8   & 1.4$^{+0.6}_{-0.4}$ & 4.2$\pm$0.8 \\
III & 2.56$\pm$0.08	  & 0.49$\pm$0.02 & 9.7$\pm$0.2  & 6.7--25     & 0.9--5.1    & 6.4--11.5 & 250$\pm$5   & 5.0$\pm$1.2	     & 5.0$\pm0.4$ \\
IV  & 1.7$^{+3.2}_{-1.3}$ & 0 (fixed)	  & $<$5.4	 & ---         & ---	     & ---	 & 280 (fixed) & 5.0 (fixed)	     & $<$11	   \\
V   & ---		  & --- 	  & --- 	 & 8.7$\pm$1.0 & 0.4$\pm$0.1 & 22$\pm$1  & 204$\pm$16  & 5.0 (fixed)	     & $<$16	   \\
VI  & 1.38$\pm$0.07	  & 0.41$\pm$0.05 & 12.9$\pm$0.4 & 5.6--10.9   & 0.8--1.9    & $\sim$11.2 & 103$\pm$28  & 0.6$\pm$0.8	      & 10$\pm$2 \\
\hline
\end{tabular}
} 
\caption{Fit results for the power spectra of groups I-VI. All
components were fit with Lorentzians. For groups I-III we fitted,
respectively, 4, 5, and 4 LF QPOs. For those observations only the
ranges of obtained fit parameters are given. Upper limits are 95\%
confidence. Rms amplitudes are integrated between 0 and $\infty$ Hz.
\label{tab:fit}} 
\end{table}

Our initial search for high frequency variability was conducted by
selecting a few groups of power spectra based on their position in
the HID with the aim to get an idea of how the variability changed
along the track in the HID and also to increase the sensitivity to
high frequency features. Six selections were made whose location in
the light curve and HID is indicated by Roman numerals I--VI. The
horizontal branch (hereafter hard branch) made up by groups I and II 
was split in two parts since power spectral properties are known to
correlate well with spectral hardness and we wanted to avoid too much
smearing of frequencies. Group III represents the transitional zone
discussed above, which stands out clearly in both the light curve and
HID. Group IV encompasses all the  observations in the vertical soft
branch and groups V and VI represent the last two observations. We
note that, apart from the one between groups I and II, all the group
boundaries in the HID coincide with considerable changes in the shape
of the low frequency part of the power spectrum.

\begin{figure}[p]
\centerline{\includegraphics[height=18cm]{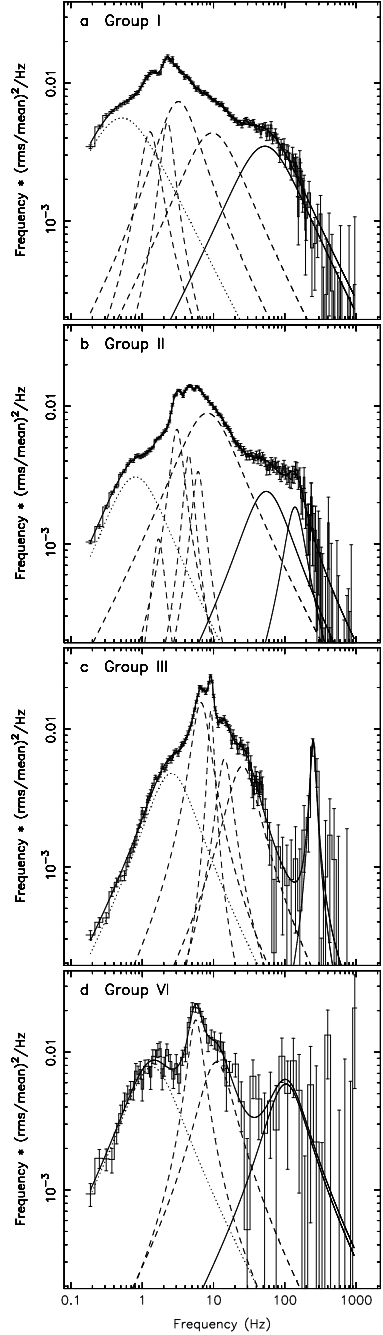}} 
\caption{The averaged 7.5--22.2 keV power spectra of groups I--III
and VI in a $\nu P(\nu)$ representation. The thick solid line shows
the best fit to the data. The number of Lorentzians used for each
group was: 6 (I), 8 (II), 6 (III), and 4 (VI). The individual
components are shown as dotted lines (noise component), dashed lines
(low frequency QPOs), and thin solid lines (high frequency QPOs). In
all cases the Poisson level was subtracted. Note that for display
purposes additional rebinning was applied to the high frequency parts
of the power spectra in order to reduce the statistical
errors.\label{fig:pds}} \end{figure}

Figure \ref{fig:pds} shows the power spectra of groups I--III and VI
in a $\nu P(\nu)$ representation - the power spectra of groups IV and
V are not shown since at high frequencies they were dominated by
statistical errors. A summary of the fit parameters can be found in
Table \ref{tab:fit}. The first three selections all show power at
high frequencies, with the most eye-catching feature being the narrow
QPO in selection III. The Lorentzian fit to this peak gives a \numax\
of 250$\pm$5 Hz, a Q-value of 5.0$\pm$1.2, and an rms amplitude of
5.0$\pm0.4$\% (6.0$\sigma$ detection), making XTE J1650--500 the
sixth black hole candidate to show such coherent high frequency
variations. In selections I, II, VI high frequency variability is
found at lower frequencies (\numax=53--139 Hz) with a lower coherence
(Q=0.25--1.4), whereas only upper limits could be determined in
selections IV and V. Note that for the power spectrum of selection
II, two Lorentzians were needed to fit the high frequency part, one
with properties similar to that of selection I, and one that was much
narrower and at a considerably higher frequency. The properties of
the $\sim$50 Hz and 250 Hz features do not change significantly when
using the non-scaled Zhang function for the subtraction of the
Poisson level.

To determine whether the frequency of the high frequency variability
changed within our selections, we made additional selections on hard
and/or soft color. Dividing group III in two parts, based on position
in the CD (not shown), we found QPOs with frequencies of 245$\pm$5 Hz
(Q$\sim$4.9) and 270$\pm$10 Hz (Q$\sim$5.3), for the spectrally hard
and soft part, respectively. This frequency difference has a
significance of 2.2$\sigma$. Dividing the hard branch, consisting of
groups I and II, in four parts, reveals that the broad feature around
50 Hz has a frequency consistent ($\sim1\sigma$) with being constant.
The $\sim$140 Hz feature in group II seems to move from 109$\pm$9 Hz,
in the hard part of that group, to 168$\pm$15 Hz in the soft part.
Both these peaks are significantly detected ($>3\sigma$), have
$Q\sim1$, and have frequencies that are more than $3\sigma$ apart.
Although this might suggest a continuous change in frequency from the
hard part of group II to the soft part of group III, we note that the
four frequencies detected in the subselections of groups II and III
are consistent (within $3\sigma$) with frequencies that are 1/3, 1/2
and 1 times the frequency of the QPO detected in group III, i.e. 250
Hz. Unfortunately, dividing our data into smaller selections, to test
the presence of intermediate frequencies did not result in
significant detections.  The 250 Hz QPO is only significantly
($>3\sigma$) detected in a few of the 33 observations of group III.
Combining those individual observations leads to a detection with a
significance that is lower than that of the detection in the whole
group, suggesting the QPO is also present in (some of) the other
observations.

The strength of the 250 Hz QPO seems to increase with energy; rms
amplitudes of $<$0.85\%, 4.5\%, and $<$12.1\% were found in,
respectively, the 2--6.2 keV, 6.2--15.0 keV, and 15.0--60 keV bands.

The low frequency part of the power spectra also shows considerable
changes between the different groups. Except for the first
observations, all power spectra of groups I and II show low frequency
QPOs, whose frequency smoothly increases with hardness from $\sim$1
Hz to $\sim$8 Hz. The low frequency part of group III generally show
a QPO around 6 and/or 9 Hz, although in a few observations no QPO is
detected at all. In groups IV and V no low frequency QPOs are
detected - they reappear in group VI with a frequency of $\sim$5.5
Hz. In general the frequencies and Q values of the Lorentzians
increase when the spectrum softens and decrease when the spectrum
hardens, although there is clearly no one-on-one relation between
hard color and frequencies (see Figures \ref{fig:lc-hid}a and
\ref{fig:pds}). Although there is a trend for both the high and low
frequency QPOs to increase with hardness, the low number of
simultaneous detections does not allow us to tell whether they show a
tight correlation.

\section{Discussion}\label{sec:disc}

Our detection of a 250 Hz QPO makes XTE J1650--500 the sixth BHC to
show such high frequency variations. The properties like Q and rms
amplitude fall in the range observed for other sources. As was
already mentioned in Section \ref{sec:intro}, HF QPOs in black hole
systems are predominantly observed when neither the non-thermal nor
the thermal component completely dominates the energy spectrum - they
have yet to be detected in the soft or hard state. XTE J1650--500
seems to be no exception to this; a preliminary analysis of the
energy spectra shows that the 250 Hz QPO was detected when the
contribution of the thermal component in the 2--200 keV range varied
between $\sim$50\% and $\sim$70\% ($\sim$10--70\% when including the
marginal detections at $\sim$110 and $\sim$170 Hz). Interestingly,
this group of observations also bridges the gap between two major
branches in our HID, a hard branch (groups I and II) and a soft
branch (group IV). The erratic changes in count rate and hardness in
this transitional zone, as opposed to the smooth changes in the soft
and hard branches, indicate that part of the accretion flow had
become unstable. Although it is not clear from our analysis what part
of the  accretion flow this would be, our results suggest that the HF
QPOs in XTE J1650--500 are enhanced when these instabilities in the
accretion flow (with a time scale of apparently hours to days) 
occur.

In addition to the 250 Hz QPO we have also found a broad feature
around 50 Hz. This feature is present in the hard branch where its
frequency was consistent  with being constant. At the moment that the
250 Hz QPO was detected this 50 Hz  disappeared, suggesting that at
least to some extent the mechanisms behind them are connected.
Although the above might suggest that the broad 50 Hz peak had
evolved into the QPO at 250 Hz, the two subselections of group II
suggest otherwise; they show peaks at $\sim$110 and $\sim$170, which
are present simultaneously with the 50 Hz peak. The constant
frequency of this broad peak is remarkable though, especially since
the LF QPOs (and possibly  also the HF QPOs, see below) show clear
changes in their frequency. Four of the eight power spectra of XTE
J1650--500 shown in Figure 2 of \citet{katoro2002} also show broad
peaks, three around 80 Hz and one around  25 Hz, and broad peaks at
similar frequencies are also found in GX 339--4 and Cyg X-1
\citep{no2000}. It is not clear whether these broad peaks are the
same as the 50 Hz peaks we found. 

The frequency behavior of the 250 Hz QPO itself is not clear either.
There are indications that the frequency increased from $\sim$110 Hz
to $\sim$270 Hz with decreasing spectral hardness, but the four
independent frequencies that were measured ($\sim$110 Hz, $\sim$170
Hz, $\sim$245 Hz, and $\sim$270 Hz) are also consistent with being
1:2:3 harmonics of each other. In a sense this behavior is similar to
what was found in XTE J1550--564; also in that source the frequency
of the HF QPO seemed to increase with decreasing spectral hardness
\citep{howiva2001}, while it was found later by \citet{remumc2002a}
that, when combining power spectra according to the type of LF QPO,
the majority of the detected HF QPOs appeared to have frequencies
that were consistent with being 1:2:3 harmonics of each other (with
the relative strength of the harmonics being related to spectra
hardness). However, in XTE J1550--564 HF QPOs were also found at
frequencies that did not obey this harmonic relation. We also note
that in the case of XTE J1650--500 the relation between HF QPOs and
the type of LF QPO is not as strong as in XTE J1550--564, since the
power spectra of group III showed considerable changes in the
coherence and relative strength of the various low frequency
components, while the frequency and harmonic content of the HF QPO
did not change significantly.

Most models proposed to explain HF QPOs in black holes
\citep{nowabe1997,stvimo1999,cuzhch1998,psno2001,abkl2001} have in
common that the oscillations are produced in the accretion disc. Some
of them predict stable frequencies \citep{nowabe1997}, whereas others
interpret them as the black hole counterparts of the kHz QPOs in
neutron star sources, and  try to predict and explain their
frequencies with respect to those of the  LF QPOs. Unfortunately, the
limited number of detections of HF QPOs in this source does not allow
us the make strong statements about any of the models. Stable
frequencies are still possible, if one interprets the different
frequencies as harmonics of each other as suggested by
\citet{abkl2001} for the pairs of peaks in GRS 1915+105 and GRO
J1655--40. On the other hand, the different frequencies might also
reflect real changes in the frequency. As noted by other authors it
might well be possible that more than one of the above mechanism are
at work at once. The broad and possibly stable feature around 50 Hz
has not been specifically addressed by the current models.  

If we simply interpret the maximum observed frequency of 270 Hz QPO
as being due to Keplerian motion at the ISCO of a Schwarzschild black
hole, we derive a mass of 8.2 $M_\odot$. For a black hole with a
lower mass no spin is required (and the radius of the
Keplerian orbit has to be larger than that of the ISCO), whereas for
more massive black holes it is. Since the properties of the broad,
skewed Fe K$\alpha$ line detected with XMM-Newton \citep{mifawi2002}
suggest that the black hole has a near maximum angular momentum, the
mass of the black hole might well be much larger than 8.2 $M_\odot$.
Assuming a mass ratio for the companion star and the black hole of
0.1, the derived mass function for the system \citep{sazuca2002}
would then imply an inclination lower than 27$^\circ$, which is
considerably lower than the values derived for GRS 1915+105
\citep[$\sim70^\circ$][]{grcumc2001}, GRO J1655--40
\citep[$\sim70^\circ$][and references therein]{grbaor2001}, and XTE
J1550--564 \citep[$\sim72^\circ$][]{orgrva2002}. If we assume that
the non-modulated X-ray emission is isotropic, this suggests that the
mechanism responsible for the HF QPOs in black hole X-ray binaries
does not produce highly beamed modulated radiation.

\acknowledgements

JH acknowledges support from Cofin-2000 grant MM02C71842. JMM is
grateful for the support of the NSF. MK acknowledges support from the
Netherlands Organization for Scientific Research (NWO). WHGL is
grateful for support from NASA.


\end{document}